\documentclass{ws-ijmpa}
\usepackage{graphicx}

\newcommand{\gev}{\mbox{~GeV}}

\newcommand{\half}{{\textstyle\frac{1}{2}}}

\newcommand{\tvec}[1]{\mbox{\boldmath{$#1$}}}

\begin{document}

\markboth{M.~Diehl}{Distribution of partons in the transverse plane}

\catchline{}{}{}{}{}

\title{ON THE DISTRIBUTION OF PARTONS \\ IN THE TRANSVERSE PLANE}

\author{M.~DIEHL} 
\address{Deutsches Elektronen-Synchroton DESY, 22603 Hamburg, Germany}

\maketitle


\begin{abstract}
Elastic nucleon form factors constrain the spatial distribution of
quarks in the impact parameter plane.  A recent analysis found that
the average impact parameter of quarks strongly depends on their
longitudinal momentum, and obtained an estimate of the orbital angular
momentum carried by valence quarks in the proton.
\end{abstract}



\section{Generalized parton distributions}

Parton densities are among the most important sources of information
on hadron structure at the level of quarks and gluons.  The
conventional densities describe the distribution of longitudinal
momentum and helicity carried by partons in a fast-moving hadron.
Generalized parton distributions (GPDs) complement this essentially
one-dimensional picture with information on the directions transverse
to the one in which the hadron
moves.\cite{Burkardt:2002hr,Diehl:2003ny} In this talk I present main
results of a study of the interplay between transverse and
longitudinal degrees of freedom.\cite{Diehl:2004cx}

Thanks to short-distance factorization GPDs appear in exclusive
scattering processes such as deeply virtual Compton scattering
($\gamma^* p\to \gamma p$) and hard meson production.  In such
processes a finite momentum is transferred to the proton.  As a
consequence, GPDs depend on two longitudinal parton momentum fractions
and in addition on the invariant momentum transfer $t$ to the proton.
Understanding the interplay between the two longitudinal parton
momenta remains an outstanding task, and in the following I consider
the simpler case where they are equal.  The so-called skewness
parameter $\xi$ is then zero, and the distributions depend on a single
longitudinal momentum fraction $x$ and on $t$.

At small $x$ and not too large $t$, phenomenological experience from
high-energy scattering suggests a simple
parameterization\cite{Goeke:2001tz}
\begin{equation}
  \label{regge-form}
H(x,t) \sim x^{- (\alpha + \alpha' t)} 
  = x^{-\alpha} \exp[\, t \alpha' \log(1/x) \,]
\end{equation}
of a generic GPD at $\xi=0$.  Literally, this form corresponds to the
exchange of a single Regge trajectory linear in $t$, but in a broader
sense it may be taken as an effective power-law valid in a limited
range of small $x$ and $t$.  For $t=0$ such a power behavior is indeed
used in many parameterizations of conventional parton densities.  It
is well known that for sea quark and gluon distributions one finds
effective exponents $\alpha$ that are very different from the
corresponding values extracted for meson or Pomeron exchange in soft
scattering processes.\cite{Landshoff:2000mu} In addition, analyses of
$J/\Psi$ production at HERA find $\alpha'$ significantly smaller than
the corresponding value for Pomeron exchange in soft reactions like
$pp\to pp$ or $\gamma p\to \rho p$.\cite{Chekanov:2004mw} On the other
hand, the typical values $\alpha\approx 0.5$ for valence quark
distributions at low momentum scale rather closely correspond to the
leading meson trajectories in Regge phenomenology of soft
interactions.  There is presently no direct measurement of $\alpha'$
for valence quark distributions.


\section{Impact parameter}

At given longitudinal momenta one can trade the dependence of a GPD on
$t$ for the dependence on the transverse momentum transferred to the
proton.  A very intuitive picture is obtained by Fourier transforming
the GPD with respect to this transverse momentum
transfer.\cite{Burkardt:2002hr,Diehl:2002he} The conjugate variable
$\tvec{b}$, called impact parameter, gives the transverse distance of
the struck parton from the center of momentum (c.o.m.) of the proton.
The c.o.m.\ of a system is the weighted average $\sum_i x_i \tvec{b}_i
/\sum_i x_i$ of the transverse positions $\tvec{b}_i$ of its
constituents, with the weights being their individual momentum
fractions $x_i$.

\begin{figure}
\begin{center}
\includegraphics[width=0.42\textwidth,%
  bb=0 -70 289 136]{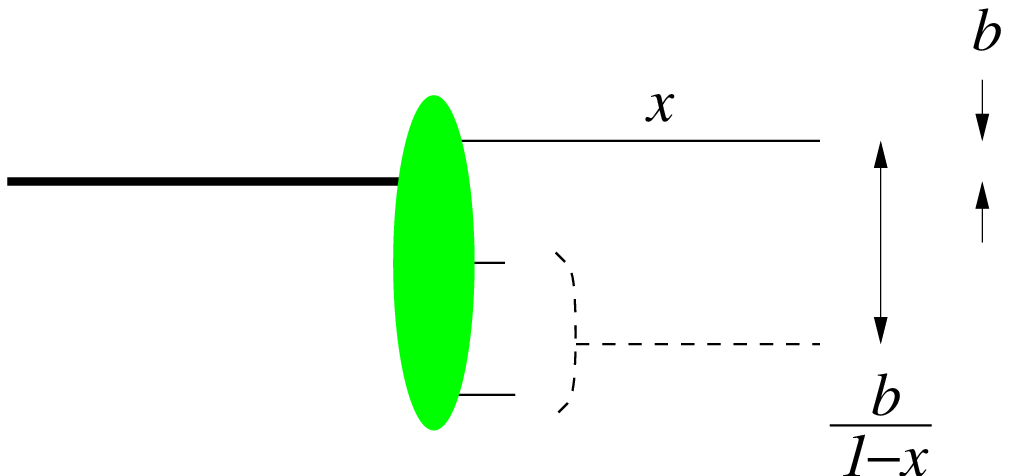}
\hfill
\includegraphics[width=0.49\textwidth,%
  bb=50 50 390 292]{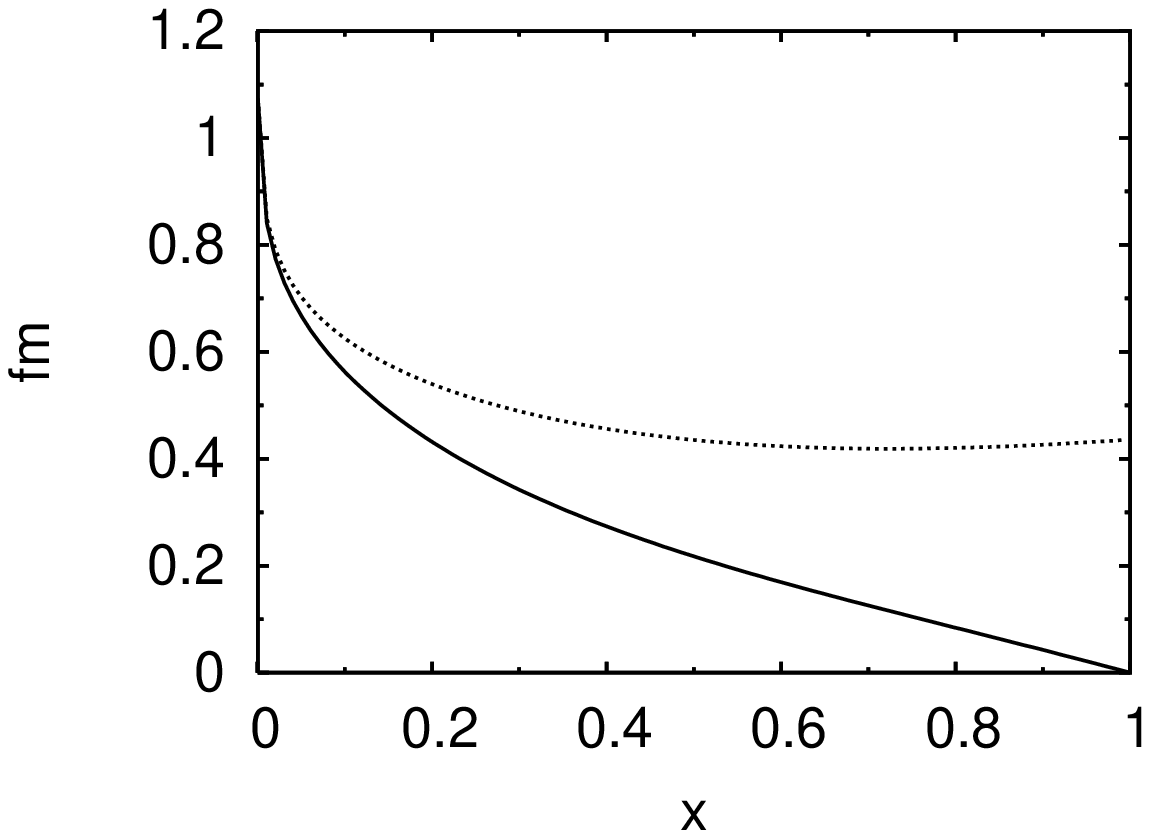}
\end{center}
\caption{\label{fig:bsquared} (a) Three-quark configuration with one
  fast quark in the proton.  The thick line denotes the c.o.m.\ of the
  proton and the dashed line the c.o.m.\ of the two spectator quarks.
  (b) Average impact parameter $[ \langle b^2 \rangle_x ]^{1/2}$ for
  valence $u$ quarks (lower curve) and the corresponding distance
  $(1-x)^{-1}\, [ \langle b^2 \rangle_x ]^{1/2}$ of the struck quark
  from the c.o.m.\ of the spectators (upper curve).}
\end{figure}

For $\xi=0$ the Fourier transformed GPDs give the probability to find
a parton with specified momentum fraction $x$ and impact parameter
$\tvec{b}$ in the proton.  The average squared impact parameter at
a given $x$ is
\begin{equation}
  \label{av-b}
\langle b^2 \rangle_x = 4 \frac{\partial}{\partial t} 
                        \log H(x,t) \Big|_{t=0} \, .
\end{equation}
This average depends on the resolution scale $\mu$ and obeys an
evolution equation similar to the one for conventional parton
densities.\cite{Diehl:2004cx} The form (\ref{regge-form}) translates
into $\langle b^2 \rangle_x$ growing like $\log(1/x)$ in the small-$x$
limit.  In the opposite limit, the impact parameter of a parton with
large $x$ tends to coincide with the c.o.m.\ of the proton (see
Fig.~\ref{fig:bsquared}a).  It is plausible to assume that, because of
confinement, the distance $b/(1-x)$ of the struck parton from the
c.o.m.\ of the remaining \emph{spectator} partons on average remains
finite for $x\to 1$.

\section{Information from the electromagnetic Dirac form factors}

Information on the valence combination $H_v^q = H^q - H^{\bar{q}}$ of
unpolarized quark and antiquark GPDs can be obtained from the sum rule
\begin{equation}
  \label{dirac-sumrule}
F_1^{p}(t) = \textstyle \int\limits_0^1 dx\, \Big[
   \frac{2}{3} H_v^{u}(x,t) 
 - \frac{1}{3} H_v^{d}(x,t) \Big]
\end{equation}
relating them to the Dirac form factor of the proton, and from the
analogous sum rule for the neutron.  We fitted the measured form
factors to an ansatz\cite{Diehl:2004cx}
\begin{equation}
  \label{H-fit}
  H_v^q(x,t) = q_v(x) \exp[\, t f_q(x) \,]
\end{equation}
with valence quark densities $q_v$ taken from the CTEQ6M
parameterization at scale $\mu=2 \gev$.  We chose $f_q(x)$ to
interpolate between limiting behaviors $\alpha' \log(1/x)$ at small
$x$ and $A^q (1-x)^2$ at large $x$, respectively corresponding to the
form (\ref{regge-form}) for $x\to 0$ and to a finite size of the
proton for $x\to 1$.  A good description of proton and neutron form
factors can be achieved, with a parameter $\alpha' \approx 0.9
\gev^{-2}$ consistent with expectations from soft Regge phenomenology.
A key result of our analysis is the strong change of the average
impact parameter over the entire $x$ range, shown in
Fig.~\ref{fig:bsquared}b.


\section{Spin and the electromagnetic Pauli form factors}

An analog of (\ref{dirac-sumrule}) relates the valence combination
$E_v^q = E^q - E^{\bar{q}}$ of proton helicity-flip GPDs to the Pauli
form factor.  Transformed to impact parameter space, a combination of
$H^q$ and $E^q$ describes the density of unpolarized quarks in a
transversely polarized proton.\cite{Burkardt:2002hr} Positivity of the
density for arbitrary quark and proton polarization implies a
bound\cite{Burkardt:2003ck}
\begin{eqnarray}
\Big[ E^q(x,t=0) \Big]^2 &\leq &  m^2 
\Big[ q(x) + \Delta q(x) \Big]\, \Big[ q(x) - \Delta q(x) \Big]\,
\nonumber \\
 && {}\times 4\, \frac{\partial}{\partial t} 
    \log \Big[H^q(x,t)\pm \tilde H^q(x,t)\Big]_{t=0}
\end{eqnarray}
in terms of the parton densities $q \pm \Delta q$ of right and left
handed quarks and the average impact parameter of quarks with either
helicity.  The strong decrease with $x$ of the right-hand side puts
strong constraints on $E^q$.  We fitted the data of proton and neutron
Pauli form factors to a parameterization
\begin{equation}
  \label{E-fit}
E_v^q(x,t) = \mathcal{N}_q\,
        x^{-\alpha} (1-x)^{\beta_q}\, \exp[\, t g_q(x) \,] ,
\end{equation}
with $g_q$ constructed as $f_q$ in (\ref{H-fit}) and normalization
factors $\mathcal{N}_q$ determined by the magnetic moments of proton
and neutron.  Good fits to the data are obtained with $\alpha \approx
0.5$, again corresponding to expectations from soft Regge
phenomenology, and for a wide range of $\beta_u$ and $\beta_d$
restricted mainly by the positivity constraints on~$E^q$.

\begin{figure}
\begin{center}
\includegraphics[width=0.48\textwidth,%
  bb=50 50 398 291]{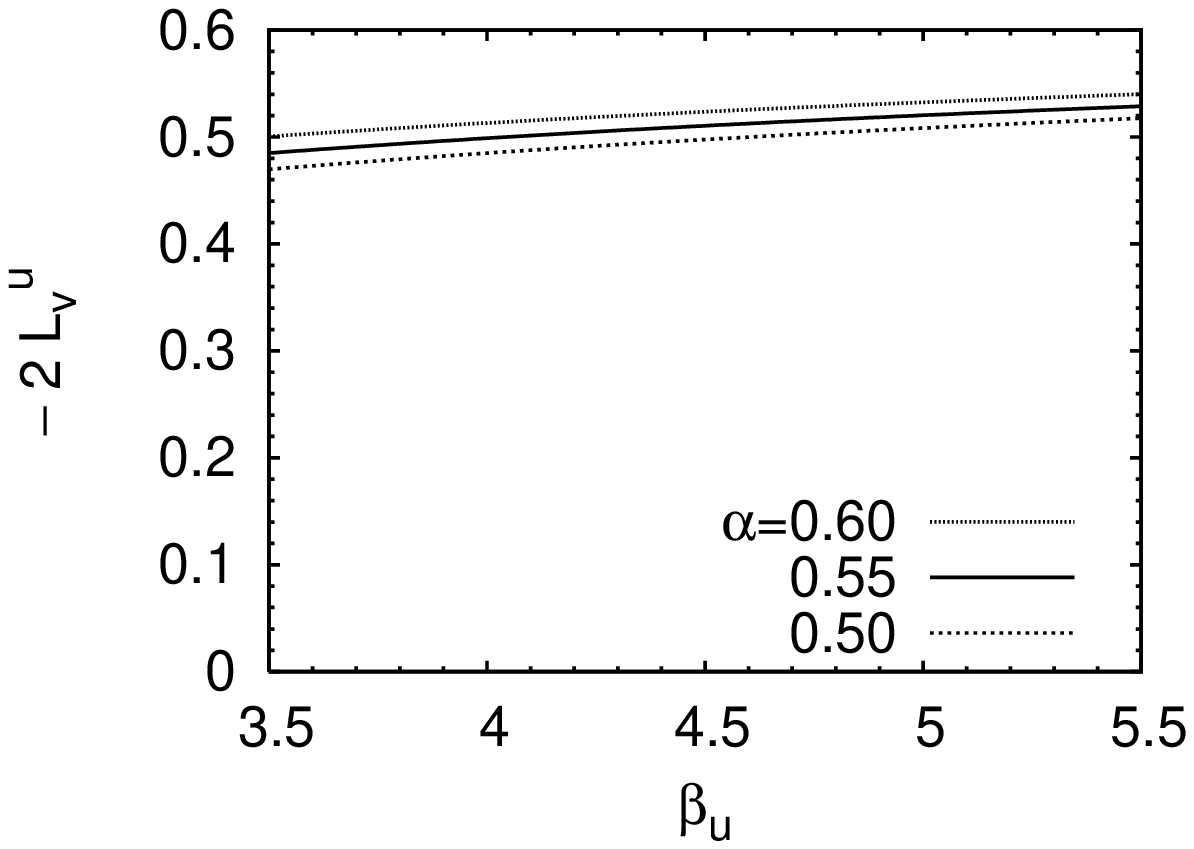}
\hfill
\includegraphics[width=0.48\textwidth,%
  bb=50 50 398 291]{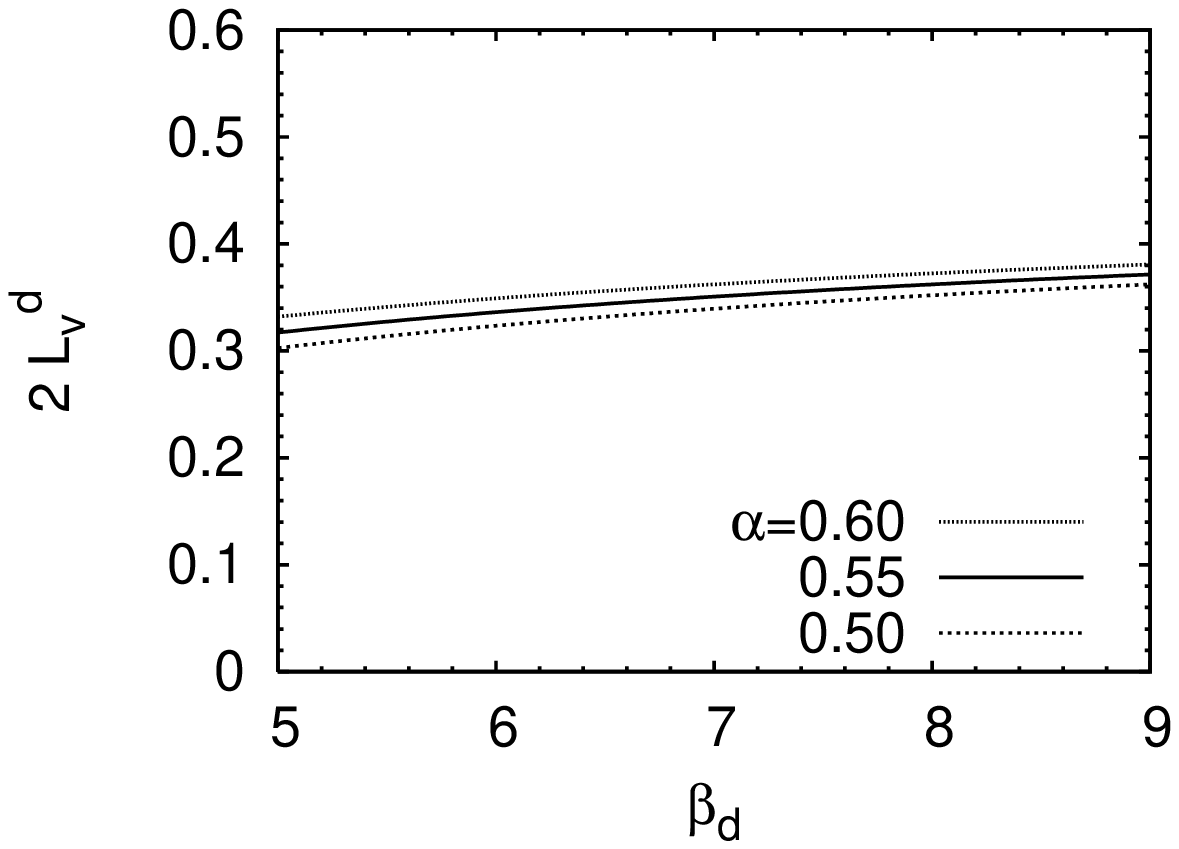}
\end{center}
\caption{\label{fig:L} The orbital angular momentum carried by valence
  quarks at scale $\mu=2 \gev$, obtained in the range of parameters
  for which we obtain a good fit to the Pauli form
  factors.\protect\cite{Diehl:2004cx}}
\end{figure}

Quite remarkably, the uncertainties on the parameters in our fit
result in very little variation of the orbital angular momentum of
valence quarks, given by Ji's sum rule as\cite{Ji:1996ek}
\begin{equation}
L_v^q =
\half \textstyle\int\limits_0^1 dx\, 
  \Big[ x E_v^q(x,t=0) + x q_v(x) - \Delta q_v(x) \Big] .
\end{equation}
The results shown in Fig.~\ref{fig:L} correspond to a range $L_v^u -
L_v^d = - \half\, ( 0.77 \div 0.92)$.  The combination $L_v^u + L_v^d
= - \half\, (0.11 \div 0.22)$ has large errors and is rather small due
to partial cancellation between the two quark flavors.

Electromagnetic form factors can only constrain the difference of
quark and antiquark distributions.  In this they are complementary to
hard scattering processes like deeply virtual Compton scattering,
which provide access to the sea quark sector.


\section*{Acknowledgments}

This work is supported by the Helmholtz Association, contract number
VH-NG-004.


\end{document}